\documentclass[article,twocolumn,showpacs,superscriptaddress,amsmath,amssymb]{revtex4-1}
\usepackage{dcolumn,epsfig,times,color,graphicx,amssymb,amsmath}% Align table columns on decimal point
 
 \definecolor{Blue}{rgb}{0,0,1}
\definecolor{NavyBlue}{rgb}{0.14,0.14,0.56}
\definecolor{rot}{cmyk}{0,1,1,0}

\begin{document}

\title{Ab initio calculations of edge-functionalized armchair graphene nanoribbons: Structural, electronic, and vibrational effects}

\author{Nils Rosenkranz}
 \email{ni-ro@physik.tu-berlin.de}
\affiliation{Institut f\"ur Festk\"orperphysik, Technische Universit\"at Berlin,
Hardenbergstr. 36, 10623 Berlin, Germany}

\author{Christian Till}
\affiliation{Institut f\"ur Festk\"orperphysik, Technische Universit\"at Berlin,
Hardenbergstr. 36, 10623 Berlin, Germany}

\author{Christian Thomsen}%
\affiliation{Institut f\"ur Festk\"orperphysik, Technische Universit\"at Berlin,
Hardenbergstr. 36, 10623 Berlin, Germany}

\author{Janina Maultzsch}
\affiliation{Institut f\"ur Festk\"orperphysik, Technische Universit\"at Berlin,
Hardenbergstr. 36, 10623 Berlin, Germany}

%\date{\today}

\begin{abstract}
We present a theoretical study on narrow armchair graphene nanoribbons (AGNRs) with hydroxyl functionalized edges. Although this kind of passivation strongly affects the structure of the ribbon, a high degree of edge functionalization proves to be particularly stable. An important consequence of the geometric deviations is a severe reduction of the band-gap of the investigated 7-AGNR. This shift follows a linear dependence on the number of added hydroxyl groups per unit cell and thus offers the prospect of a tunable band-gap by edge functionalization. We furthermore cover the behavior of characteristic phonons for the ribbon itself as well as fingerprint modes of the hydroxyl groups. A large down-shift of prominent Raman active modes allows the experimental determination of the degree of edge functionalization.

\end{abstract}

%\pacs{81.05.ue, 63.22.-m, 78.30.-j }
%63.22.Rc 	Phonons in graphene 
%73.22.Pr 	Electronic structure of graphene 
%78.67.Wj 	Optical properties of graphene (78=Raman)
% 63.22.-m Phonons or vibrational states in low-dimensional structures and nanoscale materials 
%78.30.-j 	Infrared and Raman spectra 
%81.05.ue 	Graphene 
% 73.00.00 Electronic structure and electrical properties of surfaces, interfaces, thin films, and low-dimensional structures (for electronic structure and electrical properties of superconducting films and low-dimensional structures, see 74.78.-w; for computational methodology for electronic structure calculations in condensed matter, see 71.15.-m)
% 31.15.E- Density-functional theory 
% 71.15.-m Methods of electronic structure calculations (see also 31.15.-p Calculations and mathematical techniques in atomic and molecular physics)

\maketitle

\section{Introduction} 
The unique physical properties of graphene have aroused steadily growing research efforts since 2004 equally in basic research and applications engineering [1]. Despite its exceptional electronic characteristics, the gap-less band structure of graphene impedes a direct use in many potential nanoelectronic devices. A possible way to overcome this obstacle is to design graphene nanoribbons (GNRs), which open a gap due to the lateral confinement of the electronic wave function [2,3]. The fundamental properties of these narrow stripes of a graphene sheet strongly depend on the edge structure as well as the ribbon width [3-5]. GNRs are therefore classified into armchair-edge (AGNR) and zigzag-edge ribbons (ZGNR) in analogy to the achiral types of carbon nanotubes (CNTs). Various approaches to produce GNRs have been suggested ranging from lithographic methods to the longitudinal unzipping of CNTs [6-8]. However GNRs prepared in these ways have in common one disadvantage well-known from the production of CNTs: It is hard to extract a sample of defined edge structure and width, similar to the typical chirality distributions in CNT outputs. In 2010, Cai \textit{et al.} succeeded in synthesizing one specific type of GNR with defect-free edges and a well-defined width [9]. The preparation of this 7-AGNR represents a major breakthrough in the field as the availability of predefined ribbons is crucial in order to make use of many attributes of these compounds. However, it should be noted that the reported synthesis is an exception as other GNRs so far cannot be prepared in a comparably accurate way. Typical lithographic fabrication includes treatment with \textit{e. g.} oxygen plasma which makes it very likely that GNRs are passivated other than with hydrogen. Diverse functional groups may be a side product of the manufacturing process. In this case it is vital to know how the physical properties of the GNR are affected. On the other hand, functionalization might be introduced intentionally in order to tune certain properties. Since cutting a graphene sheet into ribbons leaves dangling bonds, the edges hold a great potential for various chemical modifications. In order to describe pristine GNRs, the saturation with hydrogen is commonly assumed throughout the literature whereas divers configurations have been shown to be stable [10]. Beyond that other types of edge passivation open up an enormous number of ways to influence the attributes of GNRs. Previous works on this topic focus on ZGNRs, which yield edge states near the Fermi-level when passivated with hydrogen [11-14]. The electronic structure of ZGNRs is thus very sensitive to edge modifications. To the best of our knowledge, edge functionalization in AGNRs has only been considered as a side remark by Cervantes-Sodi \textit{et al.} who do not expect major impacts due to missing impurity levels in the band-gap [11]. However our results suggest that also other mechanisms like functionalization-induced strain play an important role in this regard.\\
In this paper, we discuss the effects of edge passivation with hydroxyl groups on structural, electronic, and vibrational properties of 7-AGNRs by means of \textit{ab initio} calculations. We demonstrate that passivation of the ribbon edges with OH-groups is energetically favorable over H passivation, with a high degree of functionalization being likely (Sec. \ref{structure}). We further present deviations in the geometries of functionalized ribbons and their hydrogen terminated counterparts. Section \ref{elBands} deals with the effect of hydroxyl functionalization on the electronic band structure of 7-AGNRs. We find a strong decrease of the band-gap with increasing degree of functionalization, which is of great interest with regard to possible applications. The influence of functional groups on characteristic vibrational modes is described in Sec. \ref{vibra}. Besides the characteristic Raman active phonons of GNRs we also consider modes specific to the hydroxyl groups.

\section{Computational details}
\begin{figure}[t]
\includegraphics[width=\columnwidth]{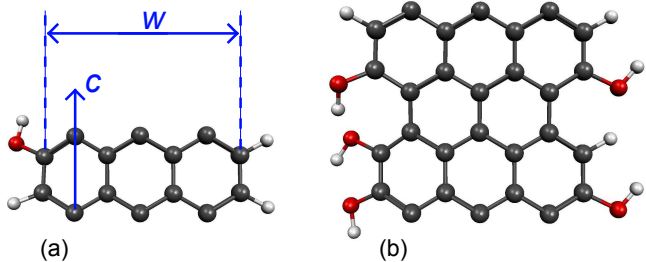}
\caption{\label{cells}
(a) Unit cell of a 7-AGNR with one hydroxyl passivated edge atom. Carbon atoms are shown in black, oxygen atoms in red (gray), and hydrogen atoms in light gray. The blue (gray) arrows indicate the lattice constant $c$ and the ribbon width $w$. (b) Super-cell of a 7-AGNR allowing more diversified configurations of edge functionalization. The example shown here has a degree of edge functionalization of 5/8.
}
\end{figure}

The 7-AGNR is an ideal model system to explore the effect of functionalization not only due to its availability in perfect quality. It also represents an appropriate compromise between growing computational cost for wide ribbons and vanishing ribbon-like characteristics for too narrow ones. We simulated 7-AGNRs with a varying number of hydroxyl groups at the edges, whereas the rest of the edge atoms was passivated with hydrogen. As we discuss exclusively chemical modifications at the edges, we define the degree of functionalization (DOF) as the number of hydroxyl groups divided by the number of available carbon edge atoms. For one unit cell the DOF can take integer multiples of 0.25 (see Fig. \ref{cells}). In order to describe intermediate values we also performed calculations of a super-cell containing two unit cells along the ribbon axis.\\
All results presented in this work were derived by means of the density functional theory package SIESTA in the local-density approximation (LDA) [15,16]. This code does not treat the core electrons explicitly but within norm-conserving pseudopotentials. Furthermore strictly localized basis sets were used to describe the valence electrons. We chose a double-$\zeta$ basis set the confinement of which was given by a maximum energy shift of 30\,meV. These two characteristics make the SIESTA method particularly efficient for low-dimensional systems like GNRs. In order to prevent interaction between neighboring images of the ribbon, the lattice vectors were chosen such that the distance between two ribbons was at least 14\,{\AA} exceeding twice the maximum radii of the basis wave functions for all elements. The integration grid in real space is ruled by an internal SIESTA parameter which was chosen fine enough to represent plane electron waves of up to 350\,Ry. Since continuous states only appear along the ribbon axis we fixed the sampling grid of the Brillouin zone to $1\times 1\times 50$. We relaxed the coordinates of all structures until the maximum atomic force was less than 0.01\,\AA. The frozen phonon approximation was used to calculate the force constant matrix. Following the procedure in previous works we scaled the $\Gamma$ point frequencies of all modes by a constant factor in order to compensate the underestimation of the bond lengths in LDA calculations [5,17,18]. The scaling factor of 0.974 found for the $E_{2g}$ mode of graphene matches the discrepancy between LDA simulations and Raman measurements of the high energy band of 7-AGNRs [9].

\section{Structure and Stability}
\label{structure}
\begin{figure}[]
\includegraphics[width=.6\columnwidth]{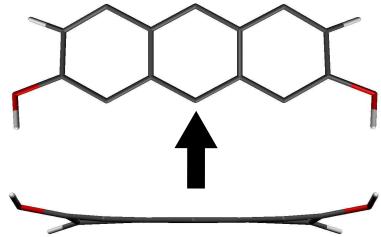}
\caption{\label{oop}
Out-of-plane arrangement of the functional groups in a 7-AGNR. The bottom picture shows the view along the ribbon axis as indicated by the black arrow.
}
\end{figure}
The addition of hydroxyl groups to the edges of 7-AGNRs induces considerable geometric effects on the ribbon itself as discussed in the following. To get a clear picture we simulated a variety of configurations with different degrees of functionalization. Apart from calculating a single unit cell of a 7-AGNR we also considered a super-cell containing two unit cells as shown for two examples in Fig. \ref{cells}. We find the planar configuration of AGNR and hydroxyl group to be metastable in contrast to findings on hydroxyl functionalized ZGNRs [11]. A bending of the functional groups out of the ribbon plane is energetically favorable by 0.12 - 0.14\,eV per group, depending on the degree of functionalization. For a single unit cell (Fig. \ref{cells}a) our results show an opposite displacement of the functional group and its neighboring edge site regardless whether the latter one is passivated with hydrogen or another OH group (see Fig. \ref{oop}). In case of the $1\times 1\times 2$ super-cell, the lower translational symmetry leads to an opposite bending of entire carbon hexagons at one edge if at least one functional group is involved.
\begin{figure}[b]
\includegraphics[width=.7\columnwidth]{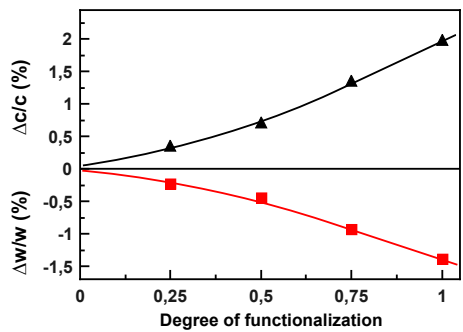}
\caption{\label{stauchung}
Relative variation of the lattice constant (black triangles) and the ribbon width (red squares) for different degrees of functionalization. Solid lines are guides to the eye.
}
\end{figure}
Apart from the out-of-plane arrangement, the edge passivation with OH groups causes a stretching along the ribbon axis accompanied by a squeezing across the width. The relative variations of the lattice constant $c$ and the ribbon width $w$ are shown in Fig.~\ref{stauchung}. Obviously the strain on the unit cell increases considerably with growing degree of functionalization. Although the total width of the ribbon is strongly decreased, the C-C bonds along the width are shortened only close to the edges. In contrast, central C-C bonds in this direction are even longer in the functionalized case than in the pristine ribbon. It turns out that the overall deformation of the unit cell of functionalized ribbons is due to a decrease of the bond angles along the width by about 2$^{\circ}$. Altogether, we observe a remarkably high Poisson ratio $\frac{\Delta w/w}{\Delta c/c}$ of $\sim 0.65$. We performed additional calculations on non-functionalized 7-AGNRs under uniaxial strain which yield a much lower Poisson ratio of 0.25 [18]. This distinction has to be taken into account when comparing functionalized and strained GNRs as discussed in Sec. \ref{elBands}.\\
Since the chemical stability of the given structures is crucial for any experimental realization, we calculated the binding energy per OH group $E_B$ for all investigated configurations as follows:
\begin{equation*}
E_B = \frac{1}{N_f}\{E(\mathrm{GNR}_{f})-[E(\mathrm{GNR}_{H-term})+N_f \cdot E(\mathrm O)]\}.
\end{equation*}

\begin{figure}[t]
\includegraphics[width=\columnwidth]{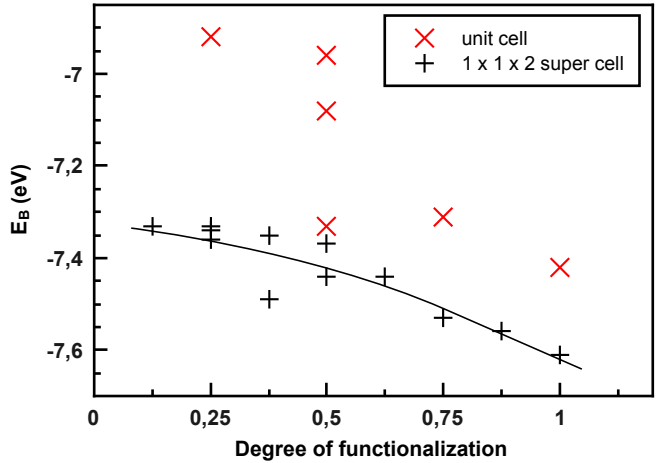}
\caption{\label{eBind}Binding energy per hydroxyl group depending on the degree of functionalization. Red diagonal crosses represent data derived from calculations of one unit cell. Black crosses are values found for a super-cell containing two unit cells along the ribbon axis. The solid black line is a guide to the eye.
}
\end{figure}

$E(\mathrm{GNR}_{f})$ and $E(\mathrm{GNR}_{H-term})$ are the total energies of a ribbon passivated with $N_f$ hydroxyl groups and a fully H-terminated ribbon, respectively. $E$(O) represents the total energy of an isolated oxygen atom. The resulting binding energies from calculations of both the super-cell and unit cell approaches are shown in Fig. \ref{eBind}. First of all we observe a substantial energy gain for the edge hydroxylation of 7-AGNRs in agreement with studies of the same effect in ZGNRs [12].
The binding energy clearly increases with growing degree of functionalization. Therefore a fully functionalized ribbon represents the most stable configuration as found for ZGNRs [12]. In many cases it is possible to construct inequivalent configurations with the same number of OH groups which then yield differing energy values. This effect is most pronounced in the unit cell approach for a degree of functionalization of 0.5. All studied cases show a clear energetic preference of single edge functionalization. Particularly high stabilities of fully saturated edges indicate the occurrence of hydrogen bonds between neighboring OH groups which have also been predicted for hydroxylized ZGNRs by Hod et al. [12]. Figure \ref{eBind} further shows consistently higher energy gains for super-cell simulations. This agrees well with the above discussed enhanced possibilities for an out-of-plane arrangement - and thus for the energy minimization - at lower translational symmetry restrictions.

\section{Electronic band structure}
\begin{figure}[t]
\includegraphics[width=\columnwidth]{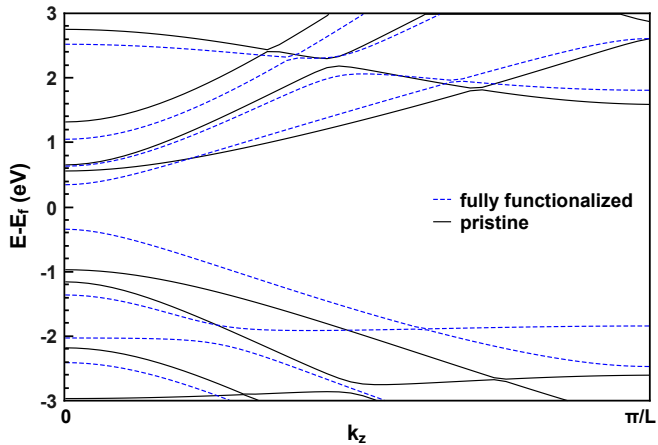}
\caption{\label{bandstructure} Band structures of an H-terminated 7-AGNR (solid black line) and a 7-AGNR with every edge atom being hydroxyl passivated (dashed blue line).}
\end{figure}

\begin{figure}[t]
\includegraphics[width=.8\columnwidth]{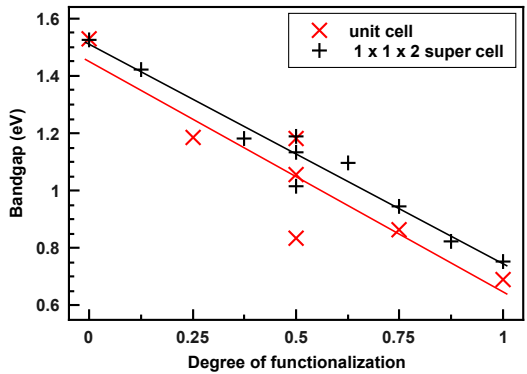}
\caption{\label{gap}Band-gap of hydroxyl-passivated 7-AGNRs depending on the degree of functionalization. Red (gray) vertical crosses and black crosses represent band-gap energies found for calculations of a unit cell and for a $1\times 1\times 2$ super-cell, respectively.}
\end{figure}

A high potential of various edge modifications for altering the band structure of ZGNRs has been reported previously [11-13]. Although AGNRs do not have edge states near the Fermi level in contrast to ZGNRs our studies reveal drastic modifications of the band structure of 7-AGNRs under edge functionalization with OH groups. Figure \ref{bandstructure} shows the electronic bands of a pristine and a maximally functionalized 7-AGNR. The most striking result is the shift of the bands close to the Fermi level which leads to a strong decrease of the band-gap. We find a linear dependence of the band-gap on the degree of functionalization as shown in Fig. \ref{gap}. Inequivalent configurations with the same degree of functionalization give different values in analogy to the discussion of the total energy in Sec. \ref{structure}. Furthermore, simulations of one unit cell yield slightly lower band-gaps than super-cell calculations. Nevertheless both approaches result in a perfectly matching linear shift of the band-gap with increasing degree of functionalization. Altogether the band-gap of 7-AGNRs can be reduced by $\sim 0.7$\,eV or almost 50\,\% by means of edge functionalization with hydroxyl groups. This is highly interesting with regard to nano-electronic applications as it offers in principle a tunable band-gap over a wide range. We did not take into account electron-electron and electron-hole interaction [19,20]. However, we believe that the general behavior will be conserved, as the dominant contribution to the band-gap shift is from geometrical effects as discussed in the following.\\
\label{elBands}
The observed band-gap dependence very much resembles previous results on uniaxially strained AGNRs, suggesting an interpretation based on the geometric deformation of the ribbon produced by the hydroxylation [18]. However, tensile strain along the ribbon axis of 2\,\%, which is the maximum deviation of the lattice constant under functionalization, corresponds to a decrease of only 0.2\,eV in band-gap. Considering the extraordinarily high Poisson ratio caused by the OH passivation of the ribbon edges, we directly compare the variations in width and band-gap. The width reduction of 1.4\,\% observed for the fully functionalized 7-AGNR (see Fig. \ref{stauchung}) would correspond to a tensile strain along the ribbon axis of 5.6\,\%. This high strain would give rise to a band-gap reduction of 0.6\,eV which comes very close to the effect found for functionalized ribbons. Based on this analysis and on the fact that AGNRs of other widths show an opposite strain dependence of the band-gap, we assume that other AGNRs show a contrary band-gap behavior under functionalization with hydroxyl groups [18]. Given the successful preparation of smaller or wider AGNRs, this would additionally offer a material with a tunable band-gap in opposite directions.

\section{Vibrational properties}
\label{vibra}
In the following we investigate the behavior of the most important vibrational modes of 7-AGNRs under different degrees of hydroxylation. We present data on the breathing-like mode (BLM) and the fundamentals of the transversal and longitudinal optical modes at $\sim 1600$\,cm$^{-1}$ which dominate the Raman spectrum [9,17,21]. In addition, characteristic modes of the hydroxyl groups and the influence of the functionalization on the stretching modes of the remaining C-H bonds are discussed.

\subsection{Breathing-like mode}
\label{BLM}

\begin{figure}[b]
\includegraphics[width=.9\columnwidth]{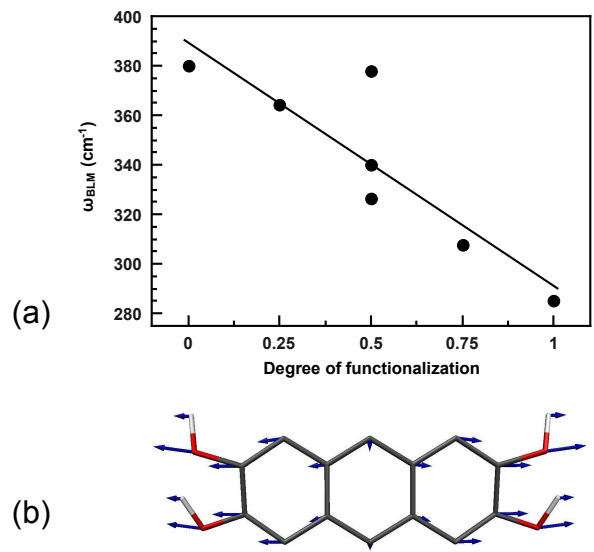}
\caption{\label{BLM-shift}
(a) Down-shift of the breathing-like mode of 7-AGNRs under functionalization with hydroxyl groups. (b) Eigenvector of the BLM of a fully hydroxylized 7-AGNR. Carbon, oxygen, and hydrogen atoms are shown in black, red (gray), and light gray, respectively. Blue arrows indicate the atomic displacements.
}
\end{figure}
Similar to the radial breathing mode in carbon nanotubes the BLM is likely to play a key role in spectroscopic characterization of GNRs. This fully symmetric vibration is highly width-sensitive but independent of the edge type. Edge-functionalization distorts the regular displacement pattern of the BLM considerably in the non-symmetric cases with one or three OH groups per unit cell. Nevertheless the basic breathing-like pattern is preserved in all investigated examples. Figure \ref{BLM-shift}(a) shows the BLM frequency of 7-AGNRs depending on the degree of functionalization. Obviously the vibration is substantially damped by the OH groups, resulting in a red-shift of 95\,cm$^{-1}$ in total. As the BLM is Raman active, this huge down-shift allows to determine the degree of functionalization experimentally. However in practice this goal may be hindered by the plenitude of Raman active modes in less symmetric compounds. The great influence of the particular arrangement of the hydroxyl groups on the stability and the band-gap is also reflected in the energy of the BLM (see Fig. \ref{BLM-shift}(a) at a degree of functionalization of 0.5). In previous work we showed that uniaxial strain also provokes a linear red-shift of the BLM. However, the dimension of the observed shift under edge passivation does not allow an exclusive ascription to the functionalization-induced strain even if we consider the unusually large Poisson ratio as described in Sec. \ref{elBands}. Most likely multiple factors contribute to the strong frequency shift. Thus, the addition of functional groups to the ribbon edges may produce a similar effect as a broadening of the ribbon which is known to reduce the BLM frequency. This assumption is supported by the displacement pattern of the BLM shown in Fig. \ref{BLM-shift}(b). All atoms of the functional groups largely follow the breathing-like motion of the ribbon itself and vibrate in phase with the carbon atoms perpendicular to the ribbon axis and in the ribbon plane. Within the picture of an expanded ribbon structure, the oxygen atoms stand for an additional dimer at each edge of the ribbon. Thus the fully functionalized 7-AGNR can be thought of as a 9-AGNR. The breathing-like mode of the hydrogen-terminated 9-AGNR is found at 300\,cm$^{-1}$. Given the rough underlying approximation this value is fairly close to the BLM frequency of 285\,cm$^{-1}$ found in the fully hydroxylated 7-AGNR.
\subsection{High-energy band}
\label{HEM}
\begin{figure}[b]
\includegraphics[width=\columnwidth]{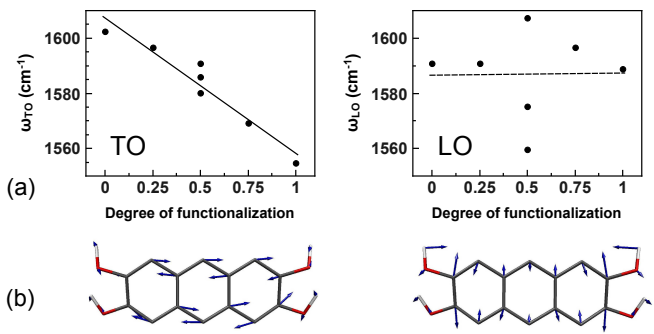}
\caption{\label{HEM-shifts} 
(a) Frequencies of the fundamentals of the LO (left) and TO phonons (right) of hydroxylized 7-AGNRs. (b) Displacement pattern of these modes. The same color code as in Fig. \ref{BLM-shift}(b) is used.
}
\end{figure}
The degeneracy of the $G$ band in graphene is lifted in GNRs, which show the longitudinal optical phonon (LO) at lower energies than the transverse mode (TO) [5]. Both phonons are clearly measurable and distinguishable in Raman experiments though under different polarization conditions [17]. As presented in Fig. \ref{HEM-shifts}(a) we find a fundamentally differing behavior of both modes under functionalization. Whereas the TO mode shifts to lower energies with increasing degree of functionalization similar to the BLM, the LO does not show such a trend. Aside from the strongly varying frequencies of the three inequivalent configurations with two OH groups per unit cell, the energy of the LO seems to be independent of the edge passivation. The large shift of the TO, in turn, is well-suited for a determination of the degree of functionalization. Like in case of the BLM a strain-based interpretation of the LO and TO behavior is problematic. As shown previously, tensile strain along the ribbon axis provokes a down-shift of both the LO and TO modes [18]. In contrast to our results on functionalized ribbons, uniaxial strain affects the LO stronger than the TO. Furthermore, even the strain-induced shift of the TO exceeds the effect of hydroxylation by a factor of 2 [18]. Once again a look at the TO and LO eigenvectors shown in Fig. \ref{HEM-shifts}(b) is helpful. A noticeable difference between the two displacement patterns concerns the behavior of the functional groups. Whereas the oscillation of the carbon atoms in the TO provokes a stretching of the C-O bonds, a strong bending of the OH group dominates the LO pattern. In both cases the motion of the outermost carbon atoms induces these characteristic vibrations of hydroxyl groups which are usually found at lower frequencies. Depending on the compound which is hydroxylized, in-plane deformations of the O-H groups are usually localized at 1260-1440\,cm$^{-1}$ and C-O stretching modes are found at 800-1150\,cm$^{-1}$ [22,23]. The coupling of the TO to the low-energetic C-O stretching mode could potentially give rise to the observed softening of the TO frequency. In contrast, the C-O-H bending which couples to the LO is expected to be closer in frequency to the LO-TO region. Therefore the LO should be affected less than the TO in agreement with our findings.

\subsection{Functionalization-specific modes}
\label{OHundCH}

\begin{figure}[b]
\includegraphics[width=.85\columnwidth]{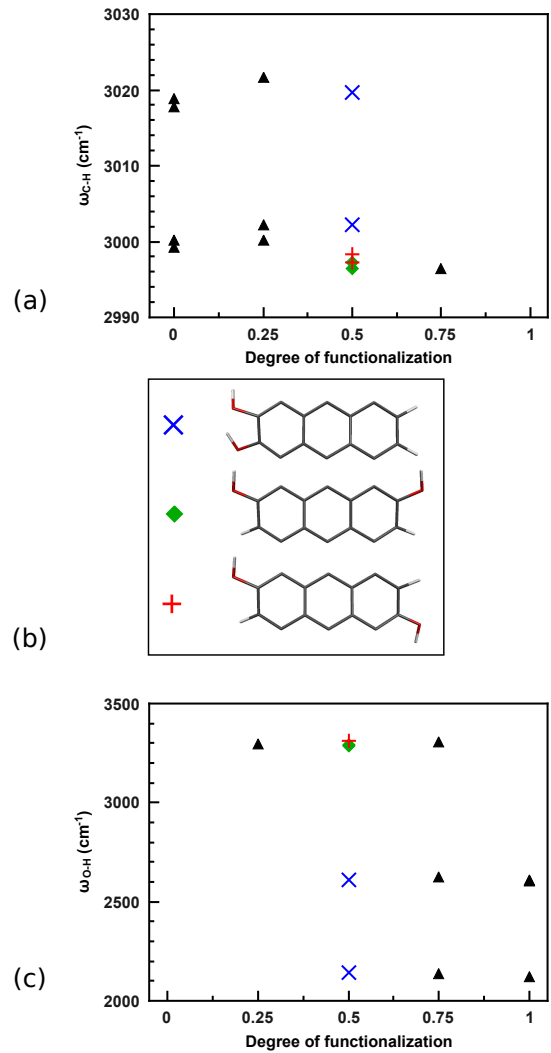}
\caption{ \label{OH-CH}
Stretching frequencies of (a) the C-H bonds and (c) the O-H bonds in 7-AGNRs depending on the degree of hydroxylation. Vibrational frequencies found for single-edge functionalization and the two different cases with half-functionalized edges are represented with the symbols stated in b). (b) Three inequivalent configurations with two OH groups per unit cell. Different elements are again displayed in the color code of Fig. \ref{BLM-shift}(b). 
}
\end{figure}

Finally we discuss the high-energy range of the vibrational spectrum comprising the stretching modes of C-H and O-H bonds. A hydrogen-terminated 7-AGNR yields two doubly degenerate pairs of C-H stretching vibrations. This number is reduced in functionalized ribbons to the number of remaining C-H bonds. Consequently we find as many O-H stretching modes as involved hydroxyl groups per unit cell. As displayed in Fig. \ref{OH-CH}(a) C-H stretching generally occurs in two narrow frequency ranges around 3000 and 3020\,cm$^{-1}$, respectively. Figure \ref{OH-CH}(c) shows three inequivalent structures with a degree of functionalization of 0.5. The two cases with one hydroxyl group per edge maintain $C_2$ rotational symmetry or, alternatively, a mirror plane. Consequently each of these configurations yields a doubly degenerate C-H stretching mode which is found 2997\,cm$^{-1}$. As expected, the non-symmetric single-edge hydroxylated 7-AGNR shows two separated C-H stretches. Measuring the C-H stretching frequencies would allow some insight in how many OH-groups are bound to the ribbon edges. From the absence of the modes at 3020\,cm$^{-1}$ a high degree of functionalization may be deduced.\\
We now turn our attention to the O-H stretching vibrations presented in Fig. \ref{OH-CH}(b). Apart from the fully functionalized 7-AGNR all cases show such modes around 3200\,cm$^{-1}$. It turns our that these are vibrations of half functionalized edges, \textit{i. e.} of OH groups which do not have direct hydroxyl neighbors. As soon as a completely hydroxylized edge is involved we observe a severe down-shift, which is a clear indication of hydrogen bonds. Opposite-phase stretching vibrations of neighboring O-H groups are found at 2605 to 2630\,cm$^{-1}$, whereas the corresponding in-phase modes are further lowered to 2120 to 2145\,cm$^{-1}$. Consequently the fully hydroxylized 7-AGNR does not show any OH-vibrations around 3200\,cm$^{-1}$. The experimental detection of the O-H stretching modes thus allows to further narrow down the degree of functionalization of a given sample. Together with the results presented in Secs. \ref{BLM} and \ref{HEM} this suggests that an extensive characterization of functionalized 7-AGNRs is feasible via vibrational spectroscopy.

\section{Conclusion}
We presented a thorough investigation of the structural, electronic, and vibrational properties of 7-AGNRs with hydroxyl functionalized edges. The passivation of every carbon edge atom with an OH group gives the most stable configuration. At the same time this maximum functionalization has the biggest effect on the electronic band structure and on the main vibrational modes. The most striking result with regard to possible applications is the large shift of the band-gap which can be attributed to strain induced by the hydroxylation. A tunable band-gap in the given range might be very useful in future nanometer-sized electronic devices. Finally we showed that the degree of edge functionalization may be determined by Raman spectroscopy. These findings are particularly exciting in the light of the successful preparation of pure 7-AGNRs. An experimental verification of our results is thus within the realms of possibility. Moreover it will be enlightening to study the impact of hydroxyl groups on AGNRs of different widths as well as the effect of other types of edge functionalization.

\section{Acknowledgements}
We thank Marcel Mohr for useful discussions. This work was supported by the DFG under grant no. MA4079/7-1.

\end{document}